\documentclass[letterpaper,prc,twocolumn,showpacs,floatfix,nofootinbib,preprintnumbers,superscriptaddress,amsmath,amssymb]{revtex4-2}
\usepackage{hyperref}
\usepackage{amsmath}
\usepackage{amsfonts}
\usepackage{physics}
\usepackage{xspace}
\usepackage{graphicx}
\usepackage[dvipsnames]{xcolor}
\usepackage{dcolumn}           
\usepackage{bm}
\usepackage{mathtools}
\usepackage{algorithm}
\usepackage{soul}
\usepackage[noend]{algpseudocode}
\usepackage{bbm}
\usepackage{comment}

\newcommand{\gras}[1]{\boldsymbol{#1}}

\newcommand{\Vq}{V_{0}^{(\rm q)}}
\newcommand{\ALR}{\rm ALR}
\newcommand{\metaM}{\hat{M}}

\begin{document}

\preprint{LLNL-JRNL-870302}

\title{Bayesian model mixing with multi-reference energy density functional }

\author{Aman Sharma}
\email{sharma25@llnl.gov}
\affiliation{Nuclear data and Theory Group, Nuclear and Chemical Science Division, Lawrence Livermore National Laboratory, California, USA 94550}

\author{Nicolas Schunck}
\email{(Corresponding author) schunk1@llnl.gov}
\affiliation{Nuclear data and Theory Group, Nuclear and Chemical Science Division, Lawrence Livermore National Laboratory, California, USA 94550}

\author{Kyle Wendt}
\email{wendt6@llnl.gov}
\affiliation{Nuclear data and Theory Group, Nuclear and Chemical Science Division, Lawrence Livermore National Laboratory, California, USA 94550}


\begin{abstract}
Reliably predicting nuclear properties across the entire chart of isotopes is 
important for applications ranging from nuclear astrophysics to superheavy 
science to nuclear technology. To this day, however, all the theoretical models 
that can scale at the level of the chart of isotopes remain semi 
phenomenological. Because they are fitted locally, their predictive power can 
vary significantly; different versions of the same theory provide 
different predictions. Bayesian model mixing takes advantage of such imperfect 
models to build a local mixture of a set of models to make improved 
predictions. Earlier attempts to use Bayesian model mixing for mass table 
calculations relied on models treated at single-reference energy density 
functional level, which fail to capture some of the correlations caused by 
configuration mixing or the restoration of broken symmetries. In this study we 
have applied Bayesian model mixing techniques within a multi-reference energy 
density functional (MR-EDF) framework. We considered predictions of 
two-particle separation energies from particle number projection or angular 
momentum projection with four different energy density functionals -- a total 
of eight different MR-EDF models. We used a hierarchical Bayesian stacking 
framework with a Dirichlet prior distribution over weights together with an 
inverse log-ratio transform to enable positive correlations between different 
models. We found that Bayesian model mixing provide significantly improved 
predictions over results from single MR-EDF calculations. 
\end{abstract}

\maketitle


\section{Introduction}
\label{sec:intro}

Most of the nuclei in the nuclear chart are unstable making them difficult to 
study experimentally in the laboratory. Therefore one has to rely more on 
theoretical models for predicting their properties. Accurate predictions of 
nuclear properties are especially relevant to improve our understanding of the 
formation of elements in the Universe \cite{schatz2022horizons,
mumpower2016impact,aprahamian2005nuclear} and the performance of nuclear 
technologies \cite{henning2024need,kolos2022current}. In spite of constant progress, the complexity of nuclear 
interactions in many-body systems and the huge computational cost involved 
prevent ab initio approaches to be applied across the entire chart of isotopes. 
Conversely, mean-field based methods can be applied to any arbitrary nucleus 
but their more phenomenological nature lead to large uncertainties away 
from stability as predictions from different parametrizations of the mean field 
differ greatly from one another \cite{sobiczewski2018detailed}. However, one 
can leverage the combined knowledge from these different models together with available 
experimental data to get better predictions. 
Statistical methods belonging to the broader field of artificial intelligence and machine learning (AI/ML) are well 
suited to this task \cite{boehnlein2022colloquium}.

Over the last decade, there have been many efforts to use AI/ML tools to 
directly predict basic nuclear properties, e.g., nuclear binding energies 
\cite{li2024atomic,yuksel2024nuclear,niu2022nuclear,sharma2022learning,
mumpower2022physically,lovell2022nuclear,niu2018nuclear}, radii 
\cite{dong2022novel}, $\beta$-decay rates \cite{niu2019predictions} or even 
fission product yields \cite{wang2019bayesian}. Since most AI/ML methods are 
data driven and data in unstable nuclei are rather sparse, such direct AI/ML 
extrapolations of observables spanning a large range of values could become 
unwieldy. A more robust approach is to use AI/ML techniques to learn 
corrections to physics models, e.g., the discrepancy between theoretical 
predictions from these physics models and experimental measurements \cite{navarroperez2022controlling,
utama2018validating,utama2017refining,utama2016nucleara}. 

The machine learning community has developed many ensemble-learning methods to 
tackle the case where several competing models have similar predictive power 
\cite{mienye2022survey,rokach2010ensemblebased}. For example, Bayesian model 
averaging (BMA) provides a natural Bayesian framework for combining model 
predictions. It has been applied in different fields of applications 
\cite{fragoso2018bayesian,hinne2020conceptual} including various nuclear 
physics problems \cite{neufcourt2020quantified,neufcourt2019neutron,
saito2024uncertainty,qiu2024bayesian}. BMA provide a weighted average of the 
given models by assigning them weights according to their posterior 
probabilities. However, such weights are global: BMA ignores the local 
variations in the prediction accuracy of different models. Bayesian model 
mixing (BMM) strategies have been suggested to account for these local effects 
\cite{phillips2021get,semposki2022interpolating,giuliani2024model,
yannotty2024model}. A recent study demonstrated BMM's effectiveness in 
predicting two-neutron separation energies \cite{kejzlar2023local}. By 
employing a Dirichlet distribution to infer model weights and assigning them 
hierarchically based on the model input space to account for variations in the 
fidelity of different models, this approach outperformed BMA in terms of 
uncertainty quantification and predictions. It also accounted for the 
correlations between the model weights for a given nucleus and from different 
parts of the nuclear chart. 

However, the use of a Dirichlet prior and the simplex constraint imposed on the 
weights also make this approach less flexible by neglecting potential positive 
correlations between the model weights for a given nucleus. The limitations of 
simplex distributions like the Dirichlet are well studied, and various 
alternative approaches have been developed to address them \cite{ongaro2020new,
greenacre2023aitchisons,aitchison1982statistical,aitchison2003statistical}. 
In addition, parametrizing model weights only as a function of proton and 
neutron numbers (input features) can lead to unreliable extrapolations 
\cite{navarroperez2022controlling}. Finally, all attempts to predict 
two-neutron separation energies using BMA and BMM have so far relied on 
combining results from different energy functionals treated at the 
single-reference energy density functional (SR-EDF) level 
\cite{neufcourt2020quantified,neufcourt2019neutron,giuliani2024model,
kejzlar2023local}. However, predictions at the SR-EDF level have limited 
accuracy and are typically unable to describe the transition from well deformed 
to (doubly) closed-shell nuclei. Increasing the predictive power of these 
methods requires working within the multi-reference energy density functional 
(MR-EDF) framework, which allows incorporating additional correlations effects 
that cannot be captured at the SR-EDF level \cite{schunck2019energy}. 
Currently, it is computationally prohibitive to restore all broken symmetries 
at once and simultaneously perform configuration mixing. BMM might provide a 
more economical method to explore the effect of restoring different symmetries: 
restoring one symmetry at a time, for example, yield a small set of different 
theoretical models that can be mixed together in a BMM framework.

In this study, we use Bayesian model mixing on two-particle separation energies 
with models based on particle-number and angular momentum projection of 
Hartree-Fock-Bogoliubov solutions with Skyrme energy density functionals. To 
ensure the robustness of extrapolations, we parametrize the BMM weights with 
several different features that have similar probability distribution in the 
training and testing region. We also investigate the use of the inverse 
log-ratio transform \cite{aitchison1982statistical} to represent model weights, 
allowing for the incorporation of positive correlations between them.  

In section \ref{sec:theory} we give a succinct summary of nuclear mass table 
calculations with MR-EDF methods. Section \ref{sec:bmm} presents our 
implementation of Bayesian model mixing and discusses to express the weights of 
the models to allow positive correlations among them. Results are presented 
in Section \ref{sec:results}.


\section{Mass Table Calculations with Projection Methods}
\label{sec:theory}

The goal of this work was to explore the possibility to capture beyond 
mean-field correlations in a statistical mixing model. As well known, such 
correlations are the main reason why there exist systematic deviations between 
theoretical predictions and experimental values of nuclear binding energies 
near closed shells (``arches'') \cite{bender2004correlation,bender2006global,
klupfel2009variations}. They are naturally incorporated in multi-reference 
energy density functional (MR-EDF) methods, which include among others symmetry 
restoration and configuration mixing, e.g., with the generator coordinate 
method \cite{bender2003selfconsistent,schunck2019energy}. While earlier studies 
pointed to the leading role of quadrupole shape mixing, we focused here on the 
restoration of broken symmetries -- particle number and angular momentum. Both 
particle number projection (PNP) and angular momentum projection (AMP) were 
based off axially-symmetric Hartree-Fock-Bogoliubov (HFB) solutions with Skyrme 
pseudopotentials.

Applying either PNP or AMP systematically lowers 
the nuclear binding energy compared to the HFB reference: $E_{\rm proj} \leq E_{\rm HFB}$. 
In this exploratory study, we did not refit the coupling constants of the 
energy density functional (EDF). Since all the EDFs considered were adjusted at 
the single-reference level, the correlation 
energy brought by the projections thus introduces a systematic bias to the theoretical prediction. For 
example, let us assume a somewhat idealistic case where we had a 
parametrization $\mathcal{P}$ of the Skyrme EDF fitted in such a way that the 
residual $x \equiv E_{\rm HFB} - E_{\rm exp.}$ (approximately) followed a 
normal distribution $x \sim \mathcal{N}(0,\sigma)$. Then, the mean of the 
``projected'' residual $x_{\rm proj} \equiv E_{\rm proj} - E_{\rm exp.}$ would 
most likely be $\bar{x}_{\rm proj} < 0$ (and, of course, the distribution may 
not be Gaussian anymore). This is a well known effect that restoring broken 
symmetries, or doing configuration mixing, {\em without refitting the 
parameters of the functional} often degrades the quality of the results even 
though the theory is more advanced. For this reason, rather than take our data 
as the nuclear binding energy, we chose instead the two-particle separation 
energy. All results will be presented for the two-neutron separation energy $S_{2n} = B(N) - B(N-2)$. It is expected that the bias introduced by the projection will largely cancel out when taking 
the difference between two neighboring nuclei.

We calculated eight mass tables corresponding to angular momentum projection 
and particle number projection with the HFB1 \cite{schunck2015error}, SkM* 
\cite{bartel1982better}, SkI3 \cite{reinhard1995nuclear}, and SLy4 
\cite{chabanat1998skyrme} parameterizations of the Skyrme functional. The SLy4, 
SkM* and SkI3 parametrizations only apply to the particle-hole channel; in the 
particle-particle channel, we used a standard zero-range, surface-volume 
density-dependent pairing force of the type $V(\gras{r}) = \Vq 
\left( 1 - \tfrac{1}{2} \tfrac{\rho(\gras{r})}{\rho_{c}} \right)$ with $\Vq$ 
the pairing strength for particle $q\equiv n,p$ and $\rho_{c}= 0.16$ fm$^{-3}$. 
Because of the zero-range of the pairing force, a cutoff of $E_{\rm cut} = 60$ 
MeV on quasiparticle energies was included when computing densities. The 
pairing strengths of protons and neutrons were fitted on the 3-point odd-even 
mass difference in $^{232}$Th and are listed in 
Table~\ref{tab:pairing_strengths}.

\begin{table}[!htb]
    \caption{Adopted pairing strengths $V_0^q$ in MeV $\cdot$ fm$^3$ for the 
             Skyrme EDFs used in this work. They were fitted to reproduce the 
             three-point odd-even mass staggering in $^{232}$Th. }
    \label{tab:pairing_strengths}
    \begin{ruledtabular}
    \begin{tabular}{ccccc}
            & SLy4 & SkM* & SkI3 \\
    \hline
    $V_0^n$ & -300.213 & -260.946 & -351.828 \\
    $V_0^p$ & -336.292& -321.045 & -376.039 \\
    \end{tabular}
    \end{ruledtabular}
\end{table}

All the nuclear mass tables used in this study were calculated with the HFBTHO 
code \cite{marevic2022axiallydeformed}. The procedure was analogue to the one 
used, e.g., in \cite{sprouse2020propagation,navarroperez2022controlling}: for 
each nucleus, we first determined the constrained HFB solution for five 
different values of the axial quadrupole moment spanning both oblate, spherical 
and prolate shapes. We then restarted the calculation without the constraint 
and retained as HFB ground-state solution the one with the lowest energy. The 
mass tables were calculated for all even even nuclei between $Z=2$ to $Z=120$ 
and the two-particle separation energy was the criterion to identify the 
driplines. Given the HFB mass table, we then apply either PNP or AMP to obtain 
the projected solution. For PNP, we used $N_{\phi} = 9$ gauge points while for 
AMP we used $N_{\beta} = 30$ rotational angles. 


\section{Bayesian model mixing}
\label{sec:bmm}

Low-energy nuclear theory is a prime example where multiple models describing 
the same physical system, each based on different assumptions and 
approximations, can have very similar predictive power. While selecting the 
best performing model is a common practice, this approach does not fully 
leverage the collective knowledge of these models. Furthermore if all models 
are incorrect, relying solely on the top performer can still lead to misleading 
predictions. In statistical learning, model sets can be categorized into three 
classes: $\mathcal{M}$-closed, $\mathcal{M}$-complete, and $\mathcal{M}$-open 
\cite{le2017bayes}. If one of the models within the set is the true data 
generating model ($M_\dagger$), such model set is considered 
$\mathcal{M}$-closed and $M_\dagger$ can be selected based on certain 
criteria or experimental evidence. If $M_\dagger$ is not in the model 
set but can be conceptualized using existing knowledge, even if practically 
inaccessible, the set is called $\mathcal{M}$-complete. Finally, a setting is 
called $\mathcal{M}$-open if $M_\dagger$ is inconceivable and all 
models in the set approach it without being able to represent it because of 
knowledge or resource limitations. The problem being studied in this paper 
falls in the $\mathcal{M}$-open class due to our lack of knowledge the 
properties of the nuclear many-body system.

Model averaging methods are well suited for $\mathcal{M}$-open problems as they 
provide predictions by combining different models and assigning them varying 
weights \cite{yao2022bayesian}. While Bayesian Model Averaging (BMA) already 
outperforms individual models, it cannot take into account the fact that 
different models can perform better in different region of the model's input 
space. Bayesian Model Mixing (BMM) strategies are much better at capturing 
local model fidelity since they assign weights that vary as a function of the 
model inputs \cite{semposki2022interpolating,giuliani2024model,
yannotty2024model}. In this paper we use a hierarchical Bayesian stacking 
method similar to what was proposed in Ref.~\cite{kejzlar2023local} with some 
modifications that we discuss below.

Let us consider a set of $K$ models denoted by $M_k$ with $k=1,2,...,K$ that
predict the same physical quantity. We assume that we have $i = 1, 2, \dots, N$ 
experimental observations $y_i$ of this quantity. In a Bayesian stacking 
framework, we seek to build a meta model $\metaM$ as a linear combination of 
the given models. This linear combination is parametrized by a vector 
$\gras{x}$ of $p$ input features, $\metaM \equiv \metaM(\gras{x})$. At 
each point $i$, we have different realizations of this input vector so that 
we can write
\begin{equation}
 y_i \simeq \metaM(\gras{x}_i) = \sum_{k=1}^{K} w_k(\gras{x}_i)M_k(\gras{x}_i)+\sigma\epsilon_i
\label{eq:BMM}
\end{equation}
where $w_k(\gras{x}_i)$ are the model weights. The observational error 
$\epsilon_i$ is assumed to follow a unit normal distribution, $\epsilon_i 
\sim \mathcal{N}(0,1)$, and $\sigma$ is a scaling parameter. Here the weights 
of each model $M_k$ obey the simplex condition 
\begin{equation}
    \sum_k w_k(\gras{x}_i) = 1, \qquad  w_k(\gras{x}_i) \geq 0
\label{eq:weights}
\end{equation}
Now the goal of the problem is to find a suitable representation for the 
input-dependent model weights $w_k(\gras{x}_i)$. 


\subsection{Dirichlet Distribution}
\label{subsec:dirichlet}

To create a locally weighted mixture of the participating models $M_k$ while 
ensuring that the simplex constraint \eqref{eq:weights} is satisfied, we model the 
weights with a Dirichlet distribution,
\begin{equation}
    p(\gras{w}|\gras{\alpha}) \sim \prod_{k=1}^{K} w_k^{\alpha_k-1}
\end{equation}
At each experimental observation $i$, the Dirichlet parameters 
$\gras{\alpha} = ( \alpha_1,\dots,\alpha_{K})$ are functions of the input features 
$\gras{x}$ through: $\ln \alpha_k(\gras{x}_i) = \gamma_k(\gras{x}_i)$. This 
dependency enables the weights to vary across different input regions, thereby 
capturing potential correlations between them at different locations 
\cite{kejzlar2023local}. In this work, we explored two methods to represent 
$\gamma_k(\gras{x})$: 
\begin{description}
    \item[Generalized linear method] In this case, we set $\gamma_k(\gras{x}) = 
    \gras{\beta}_k \cdot \gras{x}$, where $\gras{\beta}_k$ is a vector of 
    fitting parameters of length $p$. 
    \item[Gaussian process] In this case, $\gamma_k(\gras{x})$ is a Gaussian process 
    over the input features as
    \begin{equation}
    \gamma_k(\gras{x}) \sim \mathrm{GP}(\mu_k(\gras{x}),C_k(\gras{x},\gras{x}'))
    \label{eq:gp}
    \end{equation}
    The kernel $C_k(\gras{x},\gras{x}')$ represents the covariance between the 
    values of $\gamma_k$ at two different input locations, $\gras{x}$ and 
    $\gras{x}'$. The mean function $\mu_k(\gras{x})$ is the expected value of 
    $\gamma_k$ and will control the model weights far from the training data 
    \cite{kejzlar2023local}. Since our input features are nuclear properties, 
    the covariance kernel effectively captures the correlations between these 
    properties in nuclei with different neutron and proton numbers. In this 
    study, we have used constant mean functions $\mu_k$ and exponential 
    quadratic kernel,
    \begin{equation}
    C_k(\gras{x},\gras{x}') = \eta_k \prod_{n=1}^{p}  \exp\left[ \frac{-(x_{n}-x'_{n})^2}{2l_n^2} \right],
    \label{eq:cov_kernel}
    \end{equation}
    where $\eta_k$ are the variance parameter which scales the uncertainties 
    associated with the Gaussian process predictions, and $l_n$ is the 
    correlation length corresponding to the input feature $n$, which 
    control the spatial range over which the properties of the nucleus are 
    correlated.
\end{description}
In both cases, the $\gamma_k$ function effectively models the correlations 
between the weights of different models corresponding to different nuclei. 
Concurrently, the Dirichlet distribution captures the correlations between 
weights of the same nucleus. 


\subsection{Log-ratio Transformation}
\label{subsec:alr}

The model weights $w_k$ in Eq.~\eqref{eq:BMM} live in a simplex sample space 
$\mathbb{S}^K$. This is the reason why they are modeled using a Dirichlet prior
distribution, since it is specifically designed for such spaces. However, 
Dirichlet distributions also enforce negative correlations among the weights. 
In some cases, it might be more desirable for the weights to be positively 
correlated, especially when multiple models perform well in similar regions. 
Furthermore, the constraints imposed by the simplex space make it challenging 
to directly interpret the covariance structure of the weights. 

In the early 1980s J. Aitchison proposed transforming the simplex-constrained 
variables to real space $\mathbb{R}^{K-1}$ using log-ratio transformations to 
address these limitations \cite{aitchison1984log,aitchison1982statistical}. In 
this study, we employed the additive log-ratio ({\ALR}) transformation, where 
the weights are transformed as
\begin{equation}
    \omega_k = \ln\left(\frac{w_k}{w_K}\right), \qquad k=1,2,\dots,K-1
\end{equation}
where $w_K$ serves as a reference category against which ratios are defined; 
this choice is arbitrary and does not affect the analysis. The inverse {\ALR}
transformation is given by
\begin{equation}
    w_k=\begin{cases}
         \frac{1}{\sum_{k'=1}^{K-1}\big( 1 + e^{\omega_{k'}} \big)}, & \qquad k=K  \medskip\\
         \frac{e^{\omega_{k'}}}{\sum_{k'=1}^{K-1} \big(1 + e^{\omega_{k'}}\big)}, & \qquad k=1,2,..,K-1
      \end{cases}
      \label{eq:inv_alr}
\end{equation}
A multivariate normal distribution $\mathcal{N}^{K-1}(\gras{\mu},\Sigma)$ can 
induce a much richer distribution on $\mathbb{S}^K$ by exploiting the 
inverse-{\ALR} \cite{aitchison1985practical}. In this study we sampled the 
vector of transformed weights $\gras{\omega} = (\omega_1,\dots,\omega_{K-1})$ from such a 
multivariate normal distribution,
\begin{equation}
    p(\gras{\omega}|\gras{\gamma}) \sim \mathcal{N}^{K-1}(\gras{\gamma}(\gras{x}),\Sigma), \qquad k=1,2,...,K-1
\end{equation}
where the components $\gamma_k$ of the vector $\gras{\gamma}$ were computed 
using either the generalized linear method or Gaussian process as discussed in 
the previous section. The prior distribution for the covariance matrix $\Sigma$ 
was modeled with LKJ-distributed correlations \cite{lewandowski2009generating}. 
After sampling the $\omega_k$, the model weights are subsequently obtained using 
Eq. \eqref{eq:inv_alr}. 


\section{Results}
\label{sec:results}

In this section, we summarize the results of our analysis. We give the full 
details of the numerical setup employed to training the BMM before analyzing the 
global performance of the mixture model and the particular case of the Lead 
isotopic chain. We also discuss the empirical coverage probability of the 
weights and show how they vary across the nuclear chart.


\subsection{Numerical Setup}
\label{subsec:setup}

\begin{figure}[!htb]
    \centering
    \includegraphics[width=1\linewidth]{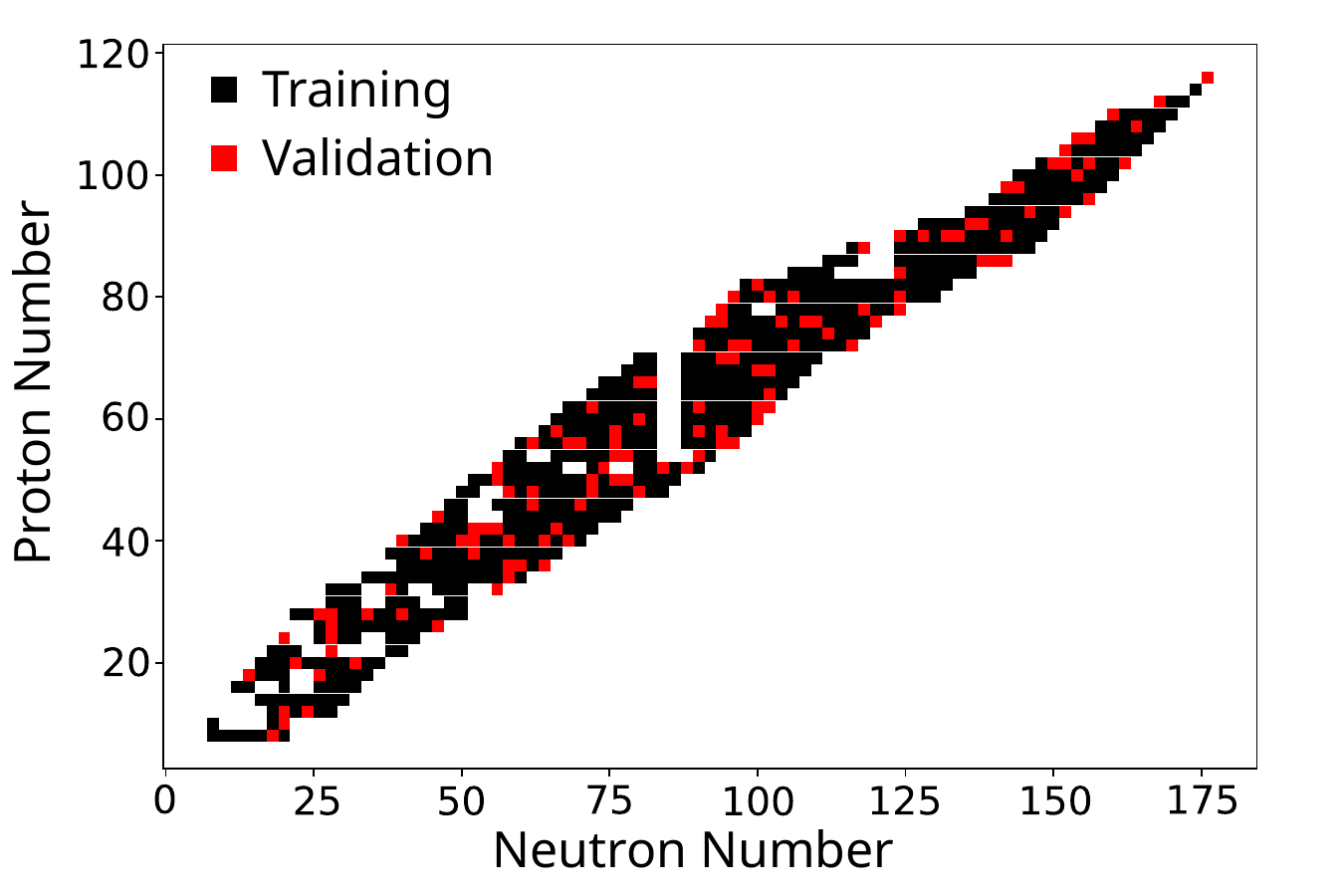}
    \caption{Location of nuclei with available experimental data included in 
             this study with the training set marked with black squares and the 
             validation set by red squares.}
    \label{fig:training}
\end{figure}

We performed BMM for two-neutron separation energies $S_{2n}$ using predictions 
from eight mass models for even-even nuclei. We used four different types of 
BMM models: the Generalised Linear Dirichlet model (GLD), the Generalised 
Linear Log-Ratio model (GLLR), the Gaussian Process Dirichlet model (GPD) and 
the Gaussian Process Log-Ratio model (GPLR). For training the models, we used 
$S_{2n}$ data from the 2020 Atomic Mass Evaluation (AME2020) \cite{wang2021ame}
for even-even nuclei with $Z\geq 8$ and $N\geq 8$.
In each mass table, MR-EDF calculations were not converged for a small subset 
of nuclei, which were excluded from further analysis. These included mostly
spherical or near-spherical nuclei near doubly closed shells, where angular
momentum projection could become numerically unwieldy. In total, $S_{2n}$ data for 624
even-even nuclei was included for both training and validation. We used a 80:20
ratio for training and validation by randomly sampling the validation set from 
the available data as presented in Fig. \ref{fig:training}.

The weights of the BMM were parametrized as a function of four input features: 
proton number, neutron number, total pairing energy and axial quadrupole 
deformation parameter $\beta_2$. The distribution of both the total pairing
energy and the $\beta_2$ deformation parameter is similar in both training and 
testing regions. Such input features should help produce more reliable 
predictions far from the training region \cite{navarroperez2022controlling}. 
For each mass model, the total pairing energy and deformation parameter values 
were obtained from the HFB solution with the corresponding energy density 
functional. Because all four input features have different range of values, 
they were normalized between 0 and 1.

\begin{figure}[!htb]
    \centering
    \includegraphics[width=1\linewidth]{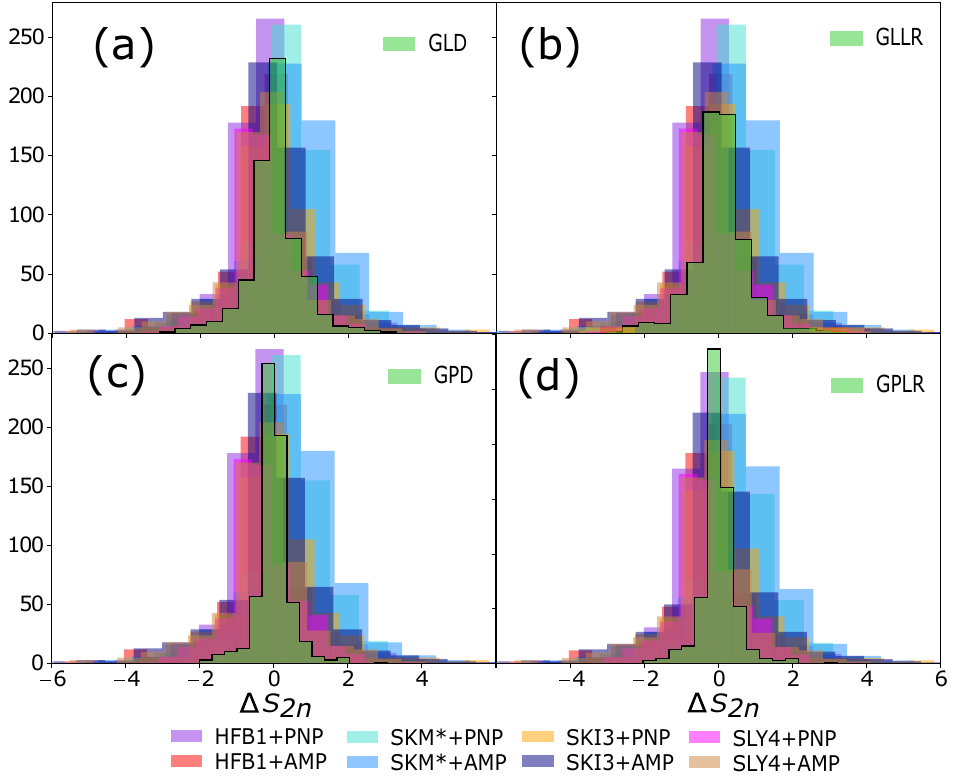}
    \caption{Histograms of the prediction error for each of the eight initial
    models and the BMM predictions, for each of the four types of BMM considered
    in this work: a) Generalized Linear Dirichlet, GLD; b) Generalized Linear
    Log-Ratio, GLLR; c) Gaussian Process Dirichlet, GPD; d) Gaussian Process
    Log-Ratio, GPLR.}
    \label{fig:predictions}
\end{figure}

The prediction errors of each mass model are heteroskedastic in nature because 
the variance differs across regions of the nuclear chart. To account for this, 
we scaled the error term in \eqref{eq:BMM} as $\sigma = \kappa/A^{2/3}$. In the 
training, all independent statistical parameters were assigned the same prior 
distributions or initial values across all variants of BMM models:
\begin{itemize}
\item For the generalized linear method, the $\beta_k$ parameters 
($\beta_k^1,\beta_k^2,...\beta_k^p$) were drawn from a unit normal distribution 
$\mathcal{N}(0,1)$.
\item The scaling factor $\kappa$ of the error was drawn from a half-normal 
prior with unit variance.
\item The parameters of the covariance kernel of Eq.~\eqref{eq:cov_kernel} of 
the Gaussian processes were drawn from an independent $\Gamma$ prior 
distribution with both shape and rate parameters set to one. This configuration 
results in a prior distribution with a mean and variance of one, which is 
equivalent to an exponential distribution.
\item The constant mean functions $\mu_k(\gras{x})$ in Eq.~\eqref{eq:gp} were 
also drawn from normal priors $\mathcal{N}(0,1)$.
\end{itemize}


\subsection{Global Performance}
\label{subsec:global}

\begin{table}[!htb]
    \caption{Root mean square (r.m.s.) deviations for each of the initial models (first eight rows) and the four variant of the mixing model (last four rows). Results are divided into training, validation and total set. Units are MeV.}
    \label{tab:rmsd_values}
    \begin{ruledtabular}
    \begin{tabular}{ccccc}
        Model.No.& Model& Training  & Validation & Total \\
        \hline
         1& HFB1+PNP & 1.136 & 1.032 & 1.116 \\
         2& HFB1+AMP & 1.084 & 1.101 & 1.087 \\
         3& SKM*+PNP & 1.229 & 1.193 & 1.222 \\
         4& SKM*+AMP & 1.534 & 1.136 & 1.463 \\
         5& SKI3+PNP & 1.202 & 1.195 & 1.201 \\
         6& SKI3+AMP & 1.333 & 1.349 & 1.336 \\
         7& SLY4+PNP & 0.977 & 1.015 & 0.985 \\
         8& SLY4+AMP & 1.184 & 1.254 & 1.198 \\
         \hline
         9& GLD      & 0.736 & 0.680 & 0.725 \\           
        10& GLLR     & 0.749 & 0.684 & 0.736 \\        
        11& GPD      & 0.440 & 0.606 & 0.478 \\
        12& GPLR     & 0.448 & 0.616 & 0.486 \\
    \end{tabular}
    \end{ruledtabular}
\end{table}

At the SR-EDF level, the average r.m.s. error on $S_{2n}$ is 1.05 MeV, with 
individual errors of 0.761 MeV (HFB1), 1.236 MeV (SkM*), 1.238 MeV (SkM*), and 
0.968 MeV (SLy4), respectively. Table \ref{tab:rmsd_values} lists the r.m.s. 
error of the eight MR-EDF models. Since we do not refit the EDFs, predictions 
for $S_{2n}$ are less accurate than those from SR-EDF calculations: the average 
r.m.s. error across all eight models has gone up to 1.20 MeV. However, the 
standard deviation has decreased, from approximately 200 keV at the SR-EDF 
level to 139 keV at the MR-EDF level. These numbers indicate that no model 
significantly outperforms the others and that the overall predictive power is 
similar. However, the histograms of Fig.~\ref{fig:predictions} also suggest 
that locally (= for a given nucleus $Z$ and $N$), the predictions can vary 
significantly between models. From a BMM perspective, this scenario can be 
helpful because stacked averages can produce better results when the 
contributing models are more dissimilar or distinct \cite{le2017bayes,
breiman1996stacked,clarke2003comparing}

Figure \ref{fig:predictions} shows that the BMM models provides consistent 
results, with prediction residuals clearly centered around zero. The r.m.s. 
errors of the BMM models are listed in the last four rows of Table 
\ref{tab:rmsd_values}. Both forms of the generalized linear model (GLD and 
GLLR) for the weights give similar results with an r.m.s. error of about 730 
keV ($\approx 40 \%$ improvement over the initial models), while both types of
Gaussian process models (GPD and GPLR) give an r.m.s. error $\approx$ 480 keV 
($\approx 60 \%$ improvement from the initial models). 

\begin{figure}[!htb]
    \centering
    \includegraphics[width=1\linewidth]{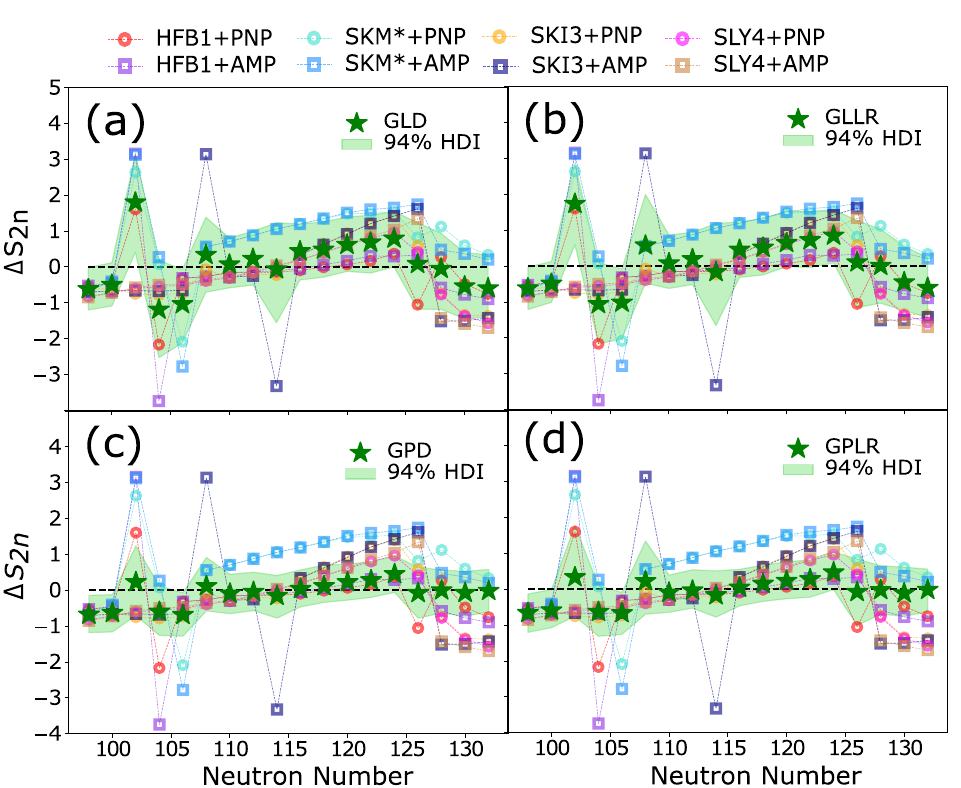}
    \caption{Prediction error of the two-neutron separation energy in Lead
    isotopes from the proton to the neutron dripline for each of the four BMM
    models: a) Generalized Linear Dirichlet, GLD; b) Generalized Linear
    Log-Ratio, GLLR; c) Gaussian Process Dirichlet, GPD; d) Gaussian Process
    Log-Ratio, GPLR.}
    \label{fig:isotopes_pb}
\end{figure}


\subsection{Example of Lead Isotopic Chain}
\label{subsec:pb}

Figure \ref{fig:isotopes_pb} illustrates the performance of different BMM 
models by showing the prediction error $\Delta S_{2n} = S_{2n}^{(\rm the.)} - 
S_{2n}^{(\rm exp.)}$ on the two-neutron results for Pb isotopes both for the 
BMM and for each of the 8 individual ones. Overall, all BMM models give similar
results with similar uncertainty bounds, whether they are based on the 
Dirichlet distribution or the log-ratio method. Because PNP or AMP are applied 
without refitting the underlying functional, $S_{2n}$ predictions often
exhibit significant deviations from experimental data, leading to abrupt peaks 
in the prediction residual plots.  BMM effectively mitigates these effects, 
especially when they are built with Gaussian processes, resulting in 
significantly improved predictions. Such Gaussian process models also reproduce 
experimental data better than Generalised linear methods because they model the 
weights as a nonlinear function of the input variables. Therefore we recommend 
using a nonlinear function to model the functional relation between weights 
and input variables. 

\begin{figure}[!htb]
    \centering
    \includegraphics[width=1\linewidth]{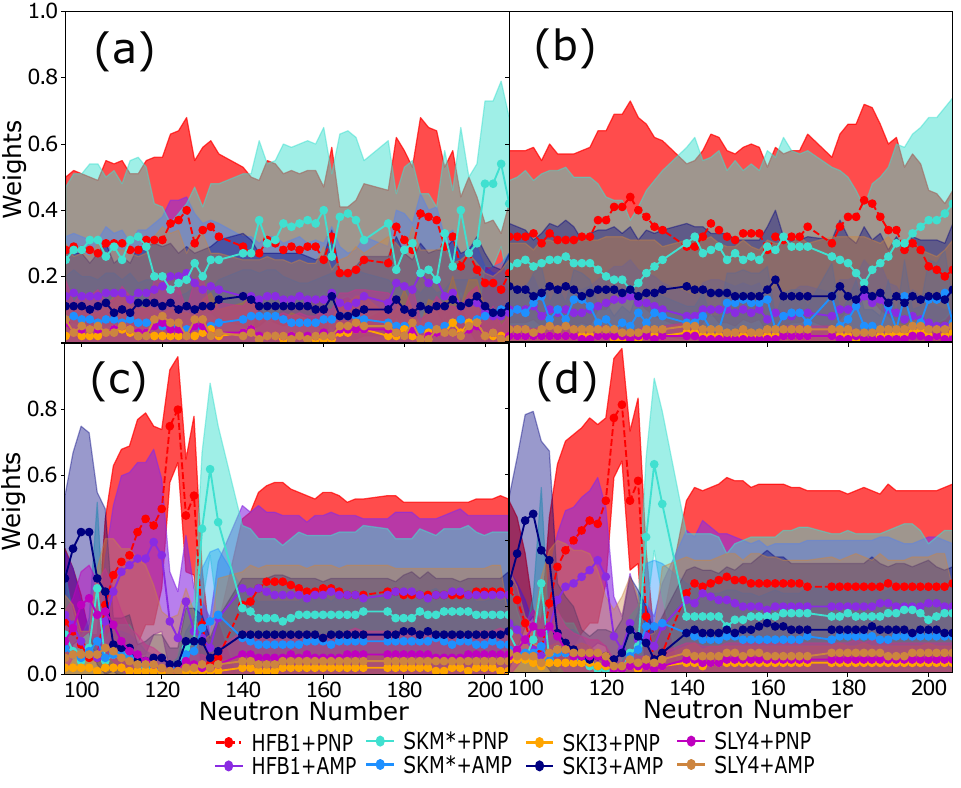}
    \caption{Posterior weights of each model in the BMM along the Lead isotopic
    line from the proton to the neutron dripline. a) Generalized Linear
    Dirichlet, GLD; b) Generalized Linear Log-Ratio, GLLR; c) Gaussian Process
    Dirichlet, GPD; d) Gaussian Process Log-Ratio, GPLR.}
    \label{fig:weights_pb}
\end{figure}

We show in Fig.~\ref{fig:weights_pb} the mean value and 1$\sigma$ uncertainty 
band of the posterior weights of each of the eight models for the full Pb 
isotopic sequence from the proton to the neutron dripline. Although the model 
weights learned using Dirichlet and log-ratio methods take similar values, 
their dependency on neutron number is different. In the case of the Dirichlet 
model -- panels (a) and (b) -- the weights vary much more rapidly as a function 
of $N$ than in the case of the GP model -- panels (c) and (d). In addition, 
these variations quickly disappear in the GP case as soon as the neutron number 
is outside the training region (at $N=132$). This is typical of the standard 
``revert to the mean'' behavior of GP-based extrapolations.

\begin{figure}[!htb]
    \centering
    \includegraphics[width=1\linewidth]{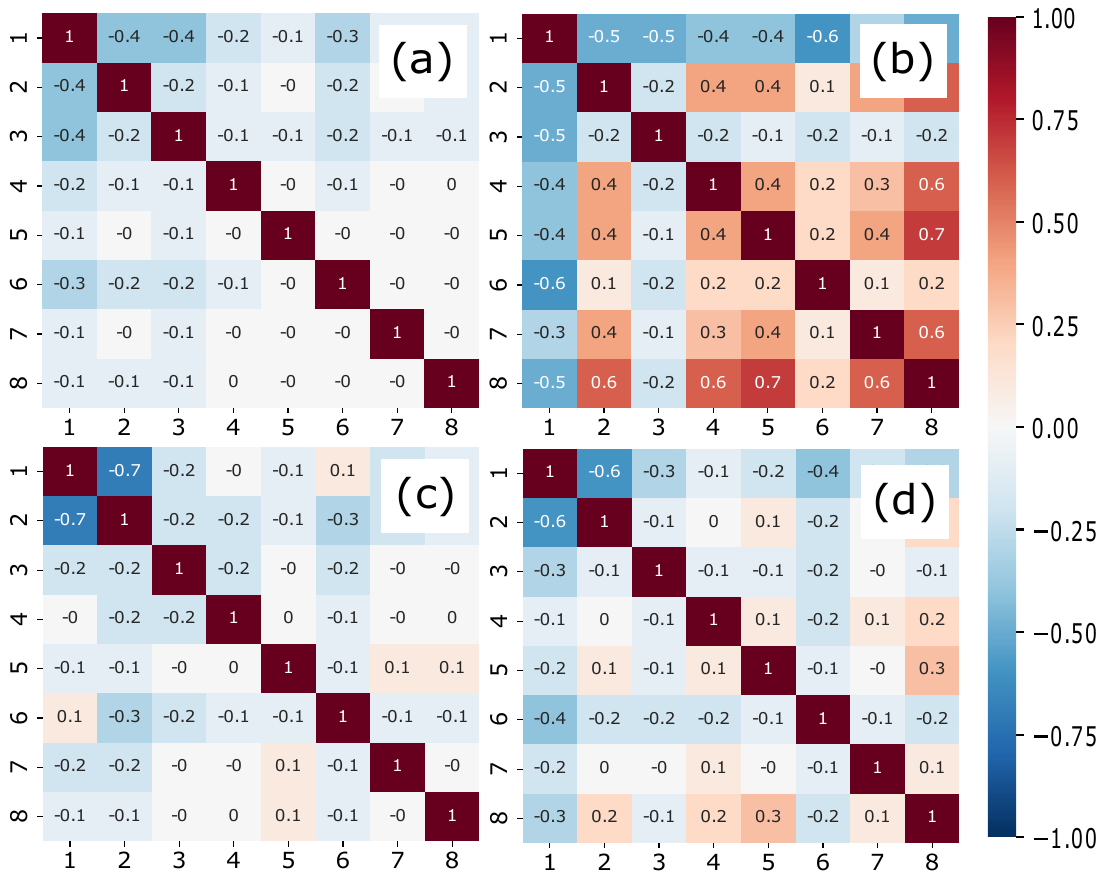}
    \caption{Correlation matrix of the eight model weights in $^{208}$Pb.
             Dirichlet models are shown in panels (a) (Generalized Linear, GLD)
             and (c) (Gausian Process, GPD); models based on the inverse log-ratio
             transformation are shown in panels (b) (Generalized Linear, GLLR)
             and (d) (Gaussian Process, GPLR).}
    \label{fig:correlations_pb}
\end{figure}

It was mentioned in Sec~\ref{subsec:alr} that Dirichlet prior distributions 
enforce negative correlations among the weights of the BMM while the {\ALR} 
also allows for positive ones. Strictly speaking, the Dirichlet distribution is 
the conjugate prior of the multinomial and categorical distributions. Since our 
likelihood follows a normal distribution, the posterior distribution of the
weights will not remain exactly Dirichlet but somewhat similar. In practice,
it manifests itself by the fact that model weights can have small positive
correlations. This is illustrated in Fig.~\ref{fig:correlations_pb} for the case
of $^{208}$Pb. The figure shows the correlation matrix between the weights of
each model for this nucleus. In
the case of the Dirichlet method -- panels (a) and (c), all off-diagonal terms 
are either negative numbers or positive numbers smaller than 0.1. In
contrast, the log-ratio method does not enforce any limitation on the sign of
the correlations between model weights and can lead to
positive correlations as high as +0.7. In fact, one observes that
weight correlations in the log-ratio method are on average stronger than in the 
Dirichlet method. This observation is true not only for the Lead isotopic chain but for
most nuclei in the nuclear chart.


\subsection{Empirical Coverage Probability}
\label{subsec:ecp}

Following \cite{kejzlar2023local}, we also computed the empirical coverage 
probability (ECP) as a function of the credibility interval to check the 
uncertainty representation of different BMM models 
\cite{gneiting2007probabilistic,gneiting2007strictly}. We recall that the ECP quantifies the
quality of the uncertainties associated with a given statistical model. For a 
credibility interval of $X \%$ -- the interval with the $X \%$ of the 
posterior distribution such that it leave equal probability density at both ends of the distribution, the value of the ECP estimates how many testing samples 
($Y \%$ in percentage) fall into this interval: if $Y>X$, the BMM model is too 
conservative; if $Y<X$, the model underestimates uncertainties. 

\begin{figure}[!htb]
    \centering
    \includegraphics[width=1\linewidth]{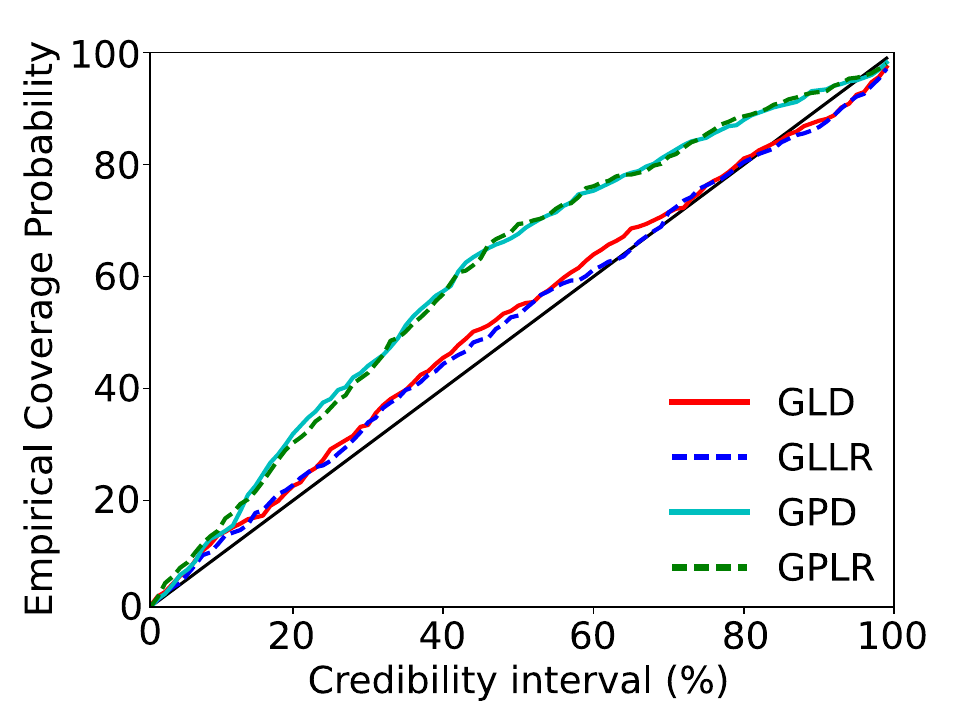}
    \caption{Empirical coverage probability as a function of credibility level
    corresponding to the four different types of Bayesian mixing models:
    Generalized Linear Dirichlet (GLD), Generalized Linear Log-Ratio (GLLR),
    Gaussian Process Dirichlet (GPD), and Gaussian Process Log-Ratio (GPLR).}
    \label{fig:coverage}
\end{figure}

Figure \ref{fig:coverage} shows the ECP for the four types of BMMs considered 
in this work. We first notice that within any given family of BMM, i.e., 
Gaussian or linear, the ECP gives similar values irrespective of whether we use 
the {\ALR} transformation or not. However, predictions corresponding to 
Gaussian process methods are significantly more conservative than with Generalized
linear methods. This could also be seen from Fig.~\ref{fig:weights_pb}: for
predictions far from the training region, the uncertainties associated with the 
value of the weights increase significantly for GPD and GPLR models.

\begin{figure*}
    \centering
    \includegraphics[width=1\linewidth]{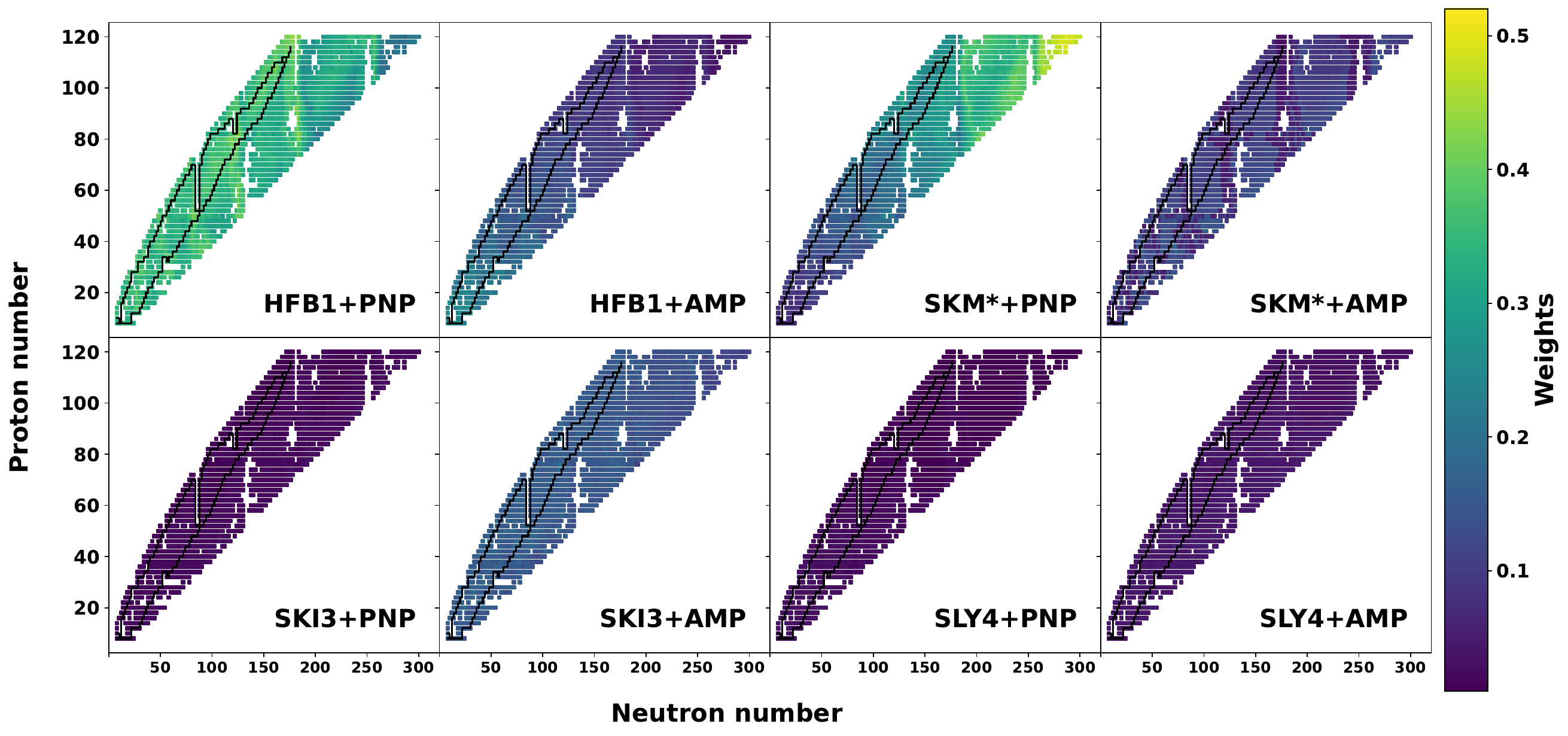}
    \caption{Posterior weight means of each of the eight models for  the generalized linear BMM model with the inverse Log-ratio transform (GLLR). }
    \label{fig:posterior_gllr}
\end{figure*}

\begin{figure*}
    \centering
    \includegraphics[width=1\linewidth]{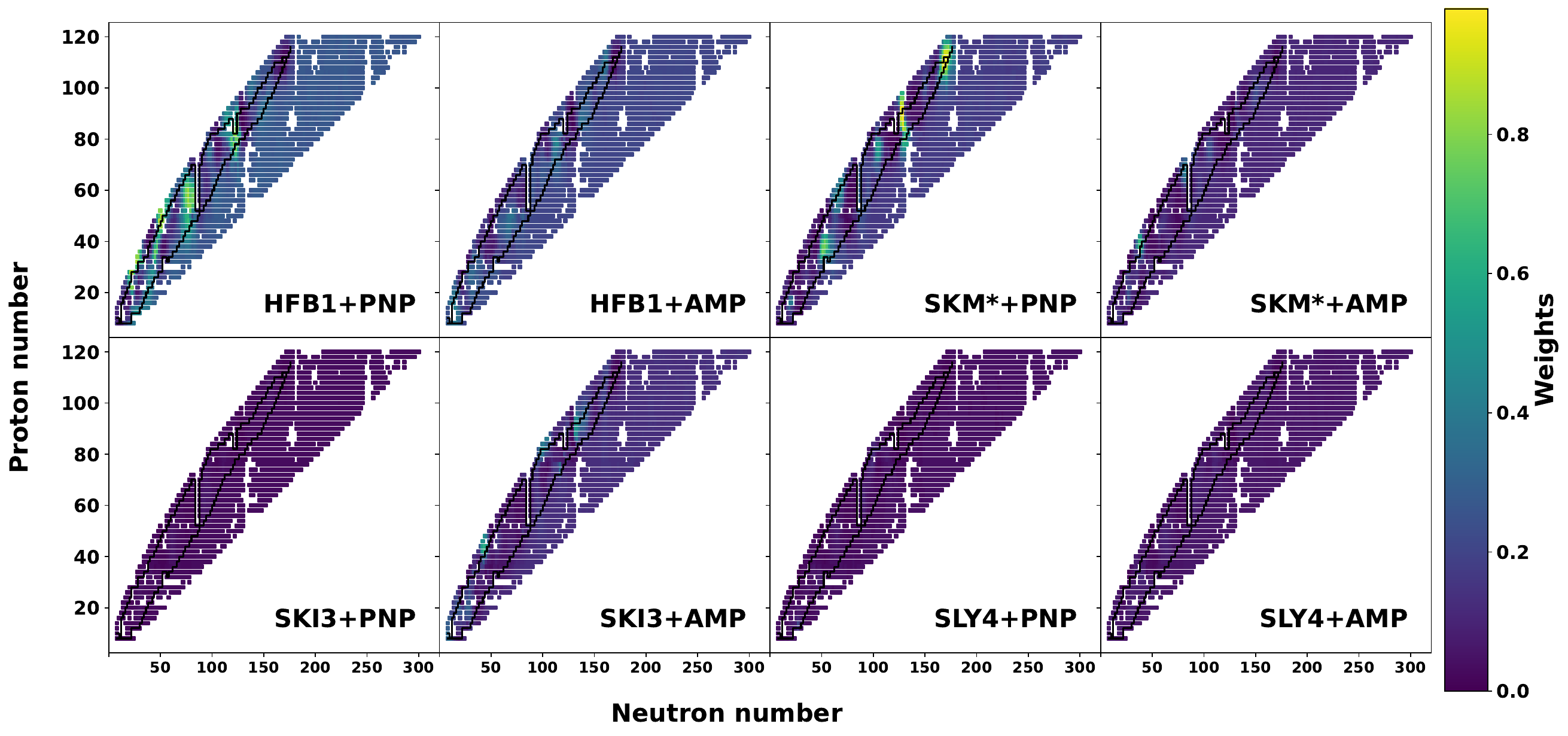}
    \caption{Same as Fig.\ref{fig:posterior_gllr} for the Gaussian process model with the inverse Log-ratio transform (GPLR).}
    \label{fig:posterior_gplr}
\end{figure*}
%


\subsection{Model Weights}
\label{subsec:weights}

We show the mean posterior weights of each model in the BMM across the entire 
nuclear chart in the case of the two BMM models with the Log-Ratio method,
which has both the lowest complexity and smallest number of degrees of freedom
and allows for correlations of arbitrary sign between model weights. Figure
\ref{fig:posterior_gllr} shows the weights for the Generalized linear model
and Fig.~\ref{fig:posterior_gplr} for the Gaussian process model.

Even though all the 8 models entering the BMM have similar predictive power, as
shown in Fig.~\ref{fig:predictions} and Table \ref{tab:rmsd_values}, some of
them clearly dominate, most notably model 1 (HFB1+PNP) and 3 (SkM*+PNP). We
note that that the GLLR method extrapolates learned trends far from the
training region. In addition, Table \ref{tab:rmsd_values} shows that the
performance of this method is in fact a little better in the validation region.
This somewhat counterintuitive result is caused by the fact that the validation
region proportionally contains more light nuclei than the training one. Because
HFB calculations are not very precise in light nuclei, the improvement brought
by the BMM is much more significant. Finally, Fig.~\ref{fig:coverage} shows
that the ECP ratio is very close to the diagonal, suggesting a near-ideal
estimate of uncertainties. Taken together, these observations suggest that GLLR
models should provide robust estimates of separation energies.

Conversely, GPLR methods perform much better in the training region than in the
validation region as shown in Table \ref{tab:rmsd_values}. At the same time,
the weights quickly converge towards a constant value as we move away from the
training set, leading into a loss of structure in each of the panels in
Fig.~\ref{fig:posterior_gplr}. These behavior are typical of overfitting and
suggest that including robust input features in the parameterization of the
weights was not sufficient to ensure reliable extrapolations. This observation
is reinforced by the much larger value of the ECP ratio.


\section{Conclusion}
\label{sec:conclusion}

In this work, we performed a Bayesian model mixing (BMM) of eight different 
mass models to make predictions of two-particle separation energies. In 
contrast to similar studies performed so far, our mass models were based on 
multi-reference energy density functional calculations, namely particle-number 
and angular momentum projection of Hartree-Fock-Bogoliubov solutions with 
Skyrme energy functionals. BMM weights were parametrized not only as a function 
of proton and neutron numbers but also as a function of the total pairing 
energy and axial quadrupole deformation in order to improve their robustness in 
extrapolations. To enhance the flexibility of the mixing model while reducing 
its complexity, we applied an inverse ALR transformation to 
model weights as function of input features. 

We found that BMM is effective in combining results from different symmetry 
restoration schemes, reducing the prediction error by a factor ranging from 
40\% to 60\% depending on the version of the BMM. Employing the inverse {\ALR} 
transformation gives similar performance as the traditional Dirichlet 
distribution method, while requiring fewer fitting parameters. It also produces 
more correlated posterior weights by allowing for positive correlations between 
them. While modeling BMM weights as a Gaussian process over input variables 
allows reducing the r.m.s. error across the entire dataset to less than 500 
keV, the uncertainties away from the training data follow the notorious
``revert to the mean'' trend of Gaussian processes, suggesting that predictions 
may not be entirely reliable. This could be mitigated by fitting the parameters 
of the energy functional concurrently with the weights of the BMM.


\begin{acknowledgments}
The authors thank Michael Grosskopf, and Henry Yuchi for enlightening discussions and 
suggestions during this work. This work was partly performed under the auspices of 
the US Department of Energy by the Lawrence Livermore National Laboratory under 
Contract DE-AC52-07NA27344. Computing support for this work came from the Lawrence 
Livermore National Laboratory Institutional Computing Grand Challenge program.
\end{acknowledgments} 

\bibliography{merged}

\end{document}